\documentstyle[times,pramana,epsf,floats]{ias}
\begin{document}
%\mark{{Gallant king...}{X Y zzz and A B zzzz}}
\title{Phenomenology of non-universal gaugino masses and implications
for the Higgs boson decay}
\author{K. Huitu,$^{1, 2}$  J. Laamanen,$^{1, 2}$, P. N. Pandita,$^{3}$ 
and Sourov Roy$^2$} 
\address{$^1$High Energy Physics Division, Department of Physical
Sciences, P.O. Box 64, FIN-00014 University of Helsinki, Finland}
\address{$^2$Helsinki Institute of Physics,
P.O. Box 64, FIN-00014 University of Helsinki, Finland}
\address{$^3$ Department of Physics, North Eastern Hill University, 
Shillong 793 022, India}
%\keywords{supersymmetry, non-universal gaugino masses, Higgs bosons}
%\pacs{12.60.Jv, 11.30.Er, 14.80.Ly}
\abstract{
Grand unified theories~(GUTs) can lead to non-universal gaugino masses 
at the unification scale.  We study the
implications of such non-universal gaugino masses for the
composition of the lightest neutralino
in  supersymmetric~(SUSY) theories based  
on  $SU(5)$ gauge group.  We also consider the phenomenological
implications of non-universal gaugino masses for the 
phenomenology of Higgs bosons in the context of Large 
Hadron Collider.}

\maketitle

\section{Introduction}
It is widely expected that some of the
supersymmetric partners of the Standard Model~(SM) particles will be 
produced at the CERN Large Hadron Collider~(LHC) which is going to start 
operation in a few years time. 
In the experimental search for supersymmetry 
the lightest SUSY particle will play a crucial role since
the heavier SUSY  particles will decay into it.
In SUSY models with R-parity conservation, the lightest SUSY
particle is absolutely stable~\cite{WIMP}. 
In most of SUSY  models the lightest
neutralino~$(\tilde\chi_1^0)$, which is an admixture of
gauginos and higgsinos, is the lightest SUSY particle~(LSP). 
Such an LSP is also a particle dark
matter candidate~\cite{DM}. From the point of view of experimental 
discovery of supersymmetry at a collider like the LHC, the LSP is 
the only SUSY particle in 
the final product of the cascade decay of a heavy SUSY particle.

In this work\cite{HLPR} we will assume
that the LSP is the lightest neutralino, and that it escapes the
collider experiments undetected. The cascade chain will typically also
contain other neutralinos~$(\tilde\chi^0_j,~ j = 2, 3, 4)$ as well as
charginos~~$(\tilde\chi^{\pm}_i,~ i = 1, 2)$. The charginos are an
admixture of charged gauginos and charged higgsinos.  The composition
and mass of the neutralinos and charginos will play a key role in the
search for SUSY  particles. The mass patterns of the neutralinos
in different SUSY models were considered in detail 
in~\cite{Pran} \cite{HLP}.

Although most of the phenomenological studies involving neutralinos
and charginos have been performed with universal gaugino masses at the
GUT scale, there is no compelling theoretical reason for
such a choice.  Thus, it is important to investigate the 
changes in the experimental signals for supersymmetry with the changes 
in the composition of neutralinos and charginos that may arise because 
of the changes in the underlying boundary conditions at the grand 
unification scale. 
In this paper we shall study the implications of the nonuniversal
gaugino masses for the phenomenology of neutral Higgs bosons in a 
$SU(5)$ supersymmetric grand unified theory.
%%%%%%%%%%%%%%%%%%%%%%%%%%%%%%%%%%%%%%%%%%%%%%%%%%%%%%%%%%%%%%%%%%%%%%5
\vskip -1.3cm
\begin{figure}[htbp]
\leavevmode
\vskip -3cm
\begin{center}
\mbox{\epsfxsize=5.truecm\epsfysize=4.truecm\epsffile{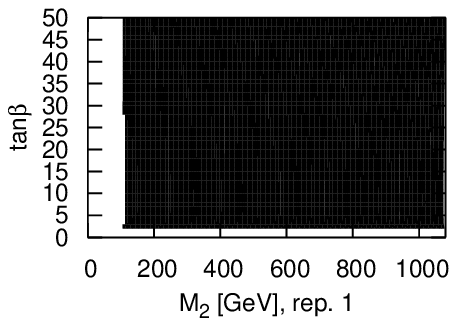}}
\mbox{\epsfxsize=5.truecm\epsfysize=4.truecm\epsffile{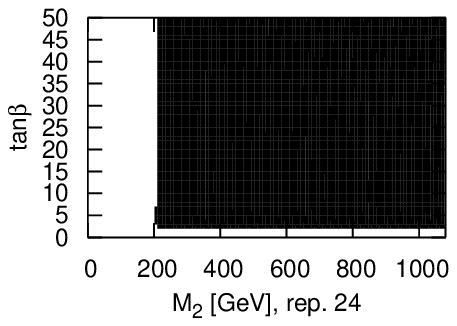}} \vskip -0.9cm
\mbox{\epsfxsize=5.truecm\epsfysize=4.truecm\epsffile{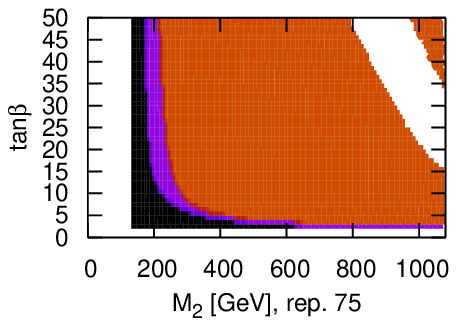}}
\mbox{\epsfxsize=5.truecm\epsfysize=4.truecm\epsffile{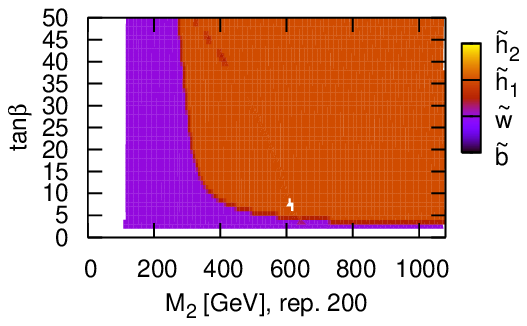}}
\end{center}
\caption{\label{composition} Main component of the lightest neutralino
in different representations of $SU(5)$ that arise in the
product~(\ref{product}).}
\vskip -1cm
\end{figure}
%%%%%%%%%%%%%%%%%%%%%%%%%%%%%%%%%%%%%%%%%%%%%%%%%%%%%%%%%%%%%%%%%%%%%%
\section{ Nonuniversal gaugino masses in supersymmetric $SU(5)$}
The masses and the compositions of neutralinos and charginos are 
determined by the soft SUSY breaking gaugino masses 
$M_i(i=1,2,3)$,  the Higgs mixing parameter $\mu$, 
and the ratio of the Higgs vacuum expectation values $tan\beta$. 
%In the simplest SUSY GUT with universal gaugino masses, $M_1 = M_2 = M_3$ 
%at the GUT scale.  
In general  
the gaugino masses need not be equal at the GUT scale.  
In SUSY  models, such as $SU(5)$ grand unified models, 
non-universal gaugino masses are generated by a non-singlet 
chiral superfield $\Phi^n$ that appears linearly in the gauge 
kinetic function $f(\Phi)$ 
which is an analytic function of the chiral superfields $\Phi$ in the 
theory \cite{CFGP}.  
When the $F$-component of $\Phi$, $F_\Phi$,  gets a VEV 
$\langle F_\Phi \rangle$, it generates the gaugino masses~($\lambda^{a,b}$ 
are gaugino fields):
\begin{eqnarray}
{\cal L}_{g.k.} \; \supset \; {{{\langle F_\Phi \rangle}_{ab}} 
\over {M_P}} \lambda^a \lambda^b +h.c. .
\end{eqnarray} 
Since the gauginos belong to the adjoint representation of $SU(5)$ with
\begin{eqnarray}
({\bf 24 \otimes 24})_{Symm} = {\bf 1 \oplus 24 \oplus 75 \oplus 200},
\label{product}
\end{eqnarray}
$\Phi$ and $F_\Phi$ can belong to, besides the singlet,
any of the non-singlet representations
{\bf 24}, {\bf 75},  and {\bf 200} of $SU(5)$. For the non-singlet
case these gaugino masses are unequal but related to one 
another~\cite{Enqvist_et_al}.  In Table~\ref{tab1}
we show the ratios of resulting gaugino masses at the GUT scale
and the electroweak scale.
\begin{table}[t] 
%\noindent
\caption{\label{tab1} Ratios of the gaugino masses at the GUT scale,
and at the electroweak scale in the 
normalization ${M_3}(GUT)$ = 1,  ${M_3}(EW)$ = 1.} 
%\centering 
\begin{tabular}{c||ccc||ccc} 
\hline \hline $F_\Phi$ & $M_1^G$ & $M_2^G$ & $M_3^G$ & $M_1^{EW}$ & $M_2^{EW}$ 
& $M_3^{EW}$ \\ \hline 
\hline {\bf 1} & 1 & 1 & 1 & 0.14 & 0.29 & 1 \\ 
{\bf 24} & -0.5 & -1.5 & 1 & -0.07 & -0.43 & 1 \\ 
{\bf 75} & -5 & 3 & 1 & -0.72 & 0.87 & 1 \\ 
{\bf 200} & 10 & 2 & 1 &1.44 & 0.58 & 1 \\ 
\hline \hline 
\end{tabular} 
\end{table} 
%%%%%%%%%%%%%%%%%%%%%%%%%%%%%%%%%%%%%%%%%%%%%%%%%%%%%%%%%%%%%%%%%%%55
In Fig.~\ref{composition} we show the dominant component of the
lightest neutralino~(LSP) for the four representations as a function
of $\tan\beta$ and $M_2$(calculated at electroweak scale) for the value 
of soft SUSY breaking scalar mass $m_0(GUT)=1$ TeV.  
For the case of the 
singlet and {\bf 24} representation, 
the dominant component is always the bino.
For the {\bf 75} dimensional representation, the situation is complicated,
and for the {\bf 200} dimensional representation 
the LSP is either a wino or a higgsino, depending on the values 
of $M_2, \mu$ and $\tan\beta$.  
%%%%%%%%%%%%%%%%%%%%%%%%%%%%%%%%%%%%%%%%%%%%%%%%%%%%%%%%%%%%%%%%%%%%%%5
%%%%%%%%%%%%%%%%%%%%%%%%%%%%%%%%%%%%%%%%%%%%%%%%%%%%%%%%%%%%%%%%%%%%%%
\section{Higgs detection using $H^0,A^0 \rightarrow
  \tilde{\chi}_2^0\tilde{\chi}_2^0 \to 4 l$}
\begin{figure}[htbp]
\begin{center}
\mbox{\epsfxsize=5.truecm\epsfysize=4.5truecm
\epsffile{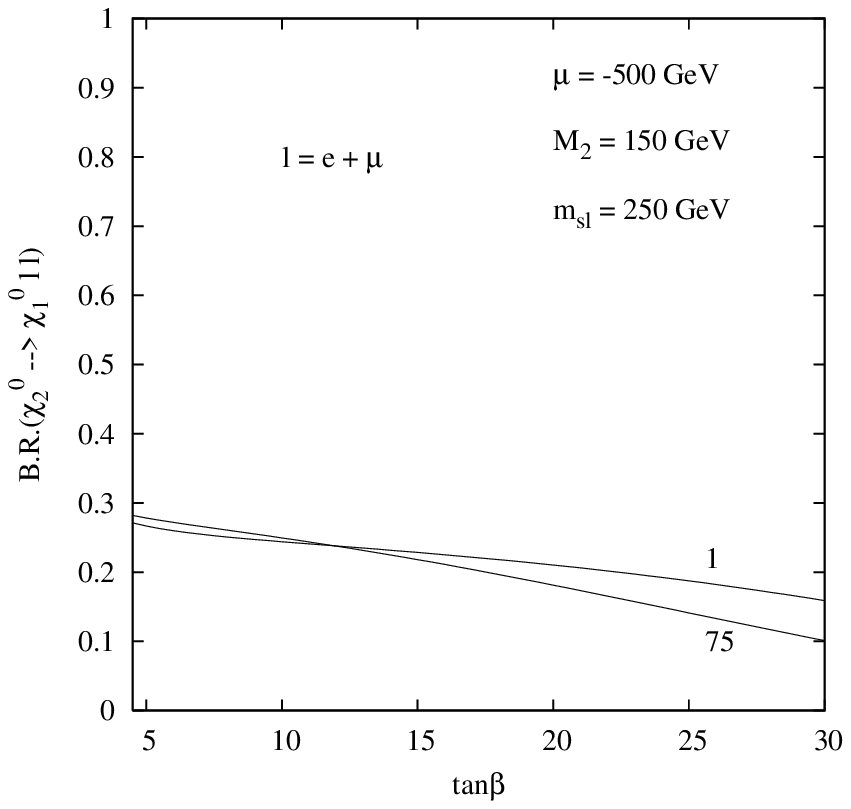}}
\mbox{\epsfxsize=5.truecm\epsfysize=4.5truecm
\epsffile{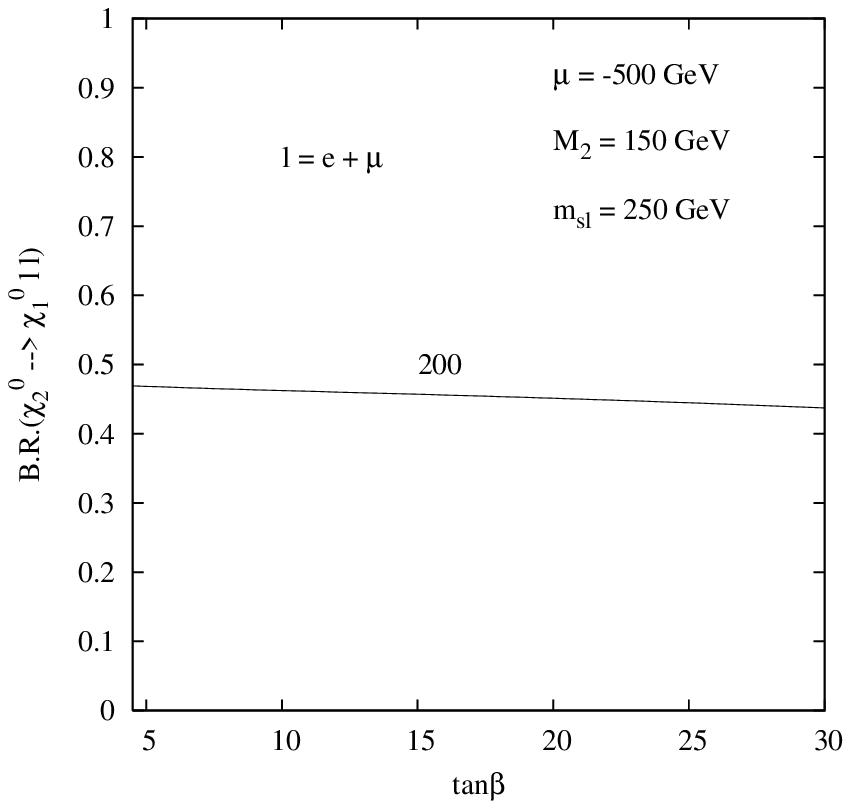}}
\end{center}
\caption{Branching ratio as a function of $\tan\beta$,
with $m_{\tilde l}$ $>$ $m_{\tilde{\chi}^0_2},$ for
the representations {\bf 1}, {\bf 75} and {\bf 200}}
\vskip -.7cm
\label{fig2}
\end{figure}

It is usually assumed that SUSY partners are too
heavy so that Higgs bosons cannot decay into SUSY particles.
However, it may well be that for the heavy Higgs bosons $H^0$, $A^0$,
and $H^\pm$ the decays to SUSY  particles are important or
even dominant. Here we will study the decay chain 
\begin{eqnarray}
H^0, A^0\to\tilde{\chi}_2^0\tilde{\chi}_2^0,\;\; \tilde{\chi}_2^0\to
\tilde{\chi}_1^0 l^+l^-,\;\; l=e,\mu,  
\label{threebody}
\end{eqnarray}
for the four different representations of $SU(5)$ in (\ref{product}).  
The decay $\tilde{\chi}_2^0\to \tilde{\chi}_1^0 l^+l^-$ depends on the 
parameters $M_2$, $M_1$, $\mu$, and $\tan\beta$, which control the 
neutralino masses and the mixing parameters, and also on the slepton 
masses $m_{\tilde l}$. As long as the direct decay of $\tilde{\chi}^0_2$ 
into $\tilde{\chi}^0_1 + Z^0$ is suppressed and the 
sleptons are heavier than the $\tilde{\chi}^0_2$, three body decays of
$\tilde{\chi}^0_2$ into charged leptons and $\tilde{\chi}^0_1$ will be
significant. In Fig.~\ref{fig2} we show a typical  branching ratio of 
the three-body decay (\ref{threebody})
as a function of $\tan\beta$ for the {\bf 1}, {\bf 75} and {\bf 200}
representations, respectively. 
In this figure the initial value of $\tan\beta$ is 4.5, since for a 
lower value of $\tan\beta$ the light Higgs mass $m_h$ is less than 
114.4 GeV, which is the LEP lower limit \cite{lephiggs}.  
We see from the figure that for higher values of $\tan\beta$ this 
branching ratio decreases since the branching ratio 
$\tilde{\chi}^0_2 \to \tilde{\chi}^0_1 \tau^+ \tau^-$ 
increases with $\tan\beta$ due to a larger Yukawa coupling.  
%%%%%%%%%%%%%%%%%%%%%%%%%%%%%%%%%%%%%%%%%%%%%%%%%%%%%%%%%%%
\vskip -0.5cm
%%%%%%%%%%%%%%%%%%%%%%%%%%%%%%%%%%%%%%%%%%%%%%%%%%%%%%%%%%%
\section{Decay of  heavy Higgs bosons into a pair of neutralinos: 
$H^0,A^0 \rightarrow \tilde{\chi}_2^0\tilde{\chi}_2^0$} 
\begin{figure}[htbp]
\vskip -1cm
\begin{center}
\mbox{\epsfxsize=5.truecm\epsfysize=5.2truecm
\epsffile{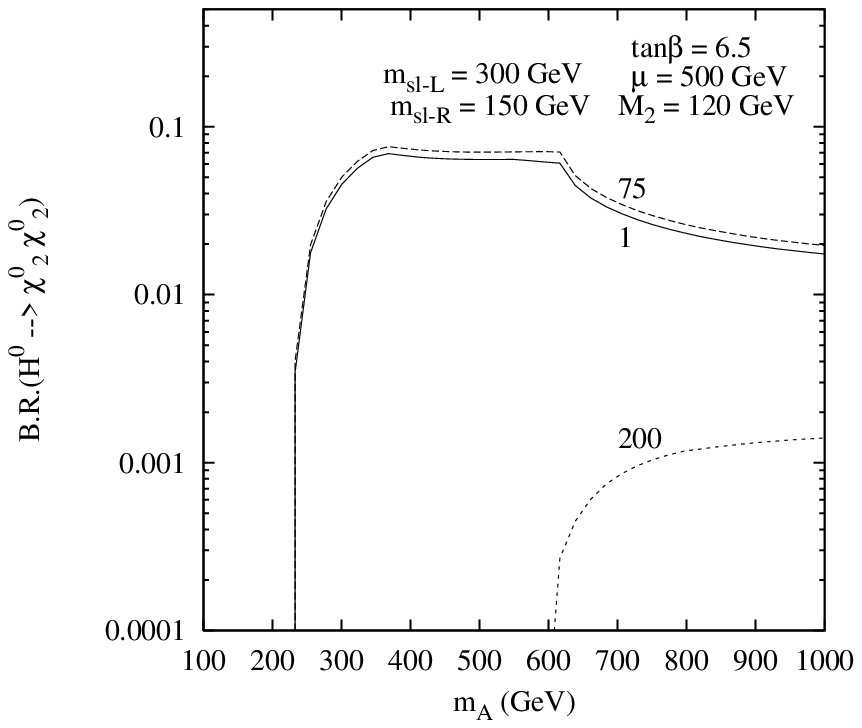}}
\mbox{\epsfxsize=5.truecm\epsfysize=5.2truecm
\epsffile{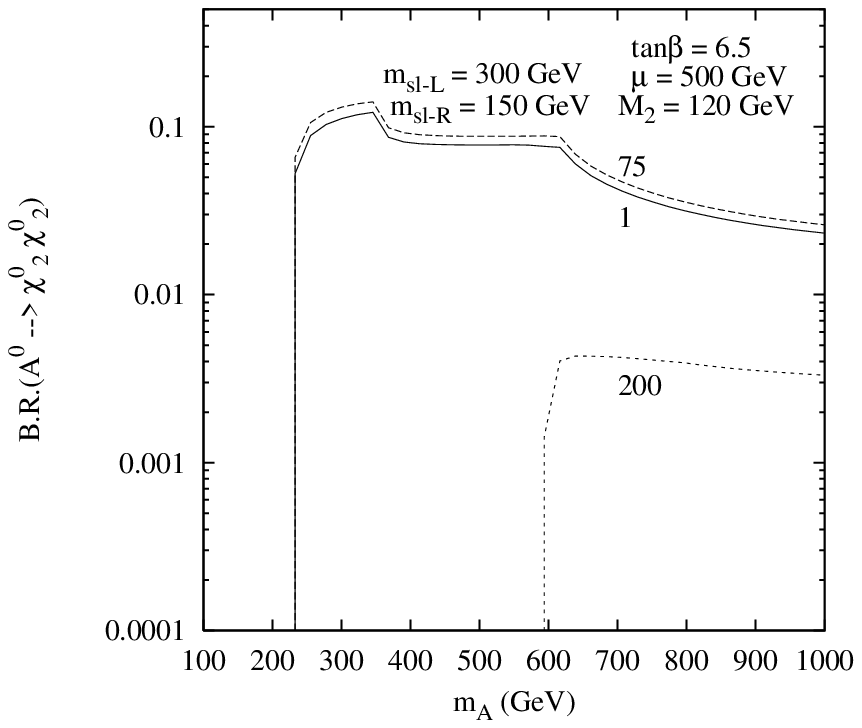}}
\end{center}
\caption{\label{fig3} The branching ratio of $H^0, A^0 \rightarrow
\tilde{\chi}_2^0\tilde{\chi}_2^0$ as a function of $m_A$ for three  
different $SU(5)$ representations in (\ref{product}). 
Here $\tan\beta$ is taken to be 6.5.}
\end{figure}

Here we study the branching ratios of the heavy Higgs bosons 
$H^0$ and $A^0$ into a pair of second lightest neutralinos. 
The decay widths and the branching ratios depend on the ratio 
of $M_1$ and $M_2$ along with other MSSM parameters. 
We have calculated the branching ratio of 
$H^0,A^0 \rightarrow \tilde{\chi}_2^0\tilde{\chi}_2^0$ 
for different $SU(5)$ representations that arise in the 
product (\ref{product}).  As an example, 
in Fig.~\ref{fig3}, we have shown the dependence of
branching ratio BR($H^0, A^0 \rightarrow \tilde{\chi}_2^0\tilde{\chi}_2^0$)
on $m_A$ for a particular choice of MSSM parameters.  
We can see that for this choice of the parameter set
and for $m_A <$ 350 GeV, the branching ratio of the decay of $A^0$ is
larger than that of the decay of the heavy Higgs scalar $H^0$ for the
representations ${\bf 1}$ and ${\bf 75}$. This is due to the fact that
for $H^0$ the total decay width is larger due to the increase in the
number of available channels to the SM particles, which leads to a
smaller branching ratio to sparticles. In the case of ${\bf 200}$
dimensional representation the threshold opens up for heavier $m_A$,
and once again the branching ratio of $A^0$ is larger than that of the
$H^0$.  The results for the representation {\bf 24} is not shown here
due to the fact that it results in the lightest neutralino mass
below the current experimental limits.
%%%%%%%%%%%%%%%%%%%%%%%%%%%%
%\vskip -1cm
%%%%%%%%%%%%%%%%%%%%%%%%%%%%%%%%%%%%%%%%%%%%%%%%%%%%%%%%%%%%%%%%%%%%%%%%%%%%
%%%%%%%%%%%%%%%%%%%%%%%%%%%%%%%%%%%%%%%%%%%%%%%%%%%%%%%%%%%%%%%%%%%%%%%%%%%

\end{document}